# Nondestructive on-chip detection of optical orbital angular momentum through a single plasmonic nanohole

Dunzhao Wei,[1] Yongmei Wang,[1] Dongmei Liu,[1] Yunzhi Zhu,[1] Weihao Zhong,[1] Xinyuan Fang,[1] Yong Zhang,[1,*] and Min Xiao[1,2,‡]

[1]National Laboratory of Solid State Microstructures, College of Engineering and Applied Sciences, and School of Physics, Nanjing University, Nanjing 210093, China

[2]Department of Physics, University of Arkansas, Fayetteville, Arkansas 72701, USA

[*] zhangyong@nju.edu.cn

[†] mxiao@uark.edu

(Dated: October 12, 2016)

**Abstract**

Optical orbital angular momentum (OAM) provides an additional dimension for photons to carry information in high-capacity optical communication. Although the practical needs have intrigued the generations of miniaturized devices to manipulate the OAM modes in various integrated platforms, the on-chip OAM detection is still challenging to match the newly-developed compact OAM emitter and OAM transmission fiber. Here, we demonstrate an ultra-compact device, i.e., a single plasmonic nanohole, to efficiently measure an optical beam's OAM state in a nondestructive way. The device size is reduced down to a few hundreds of nanometers, which can be easily fabricated and installed in the current OAM devices. It is a flexible and robust way for in-situ OAM monitoring and detection in optical fiber networks and long-distance optical communication systems. With proper optimization of the nanohole parameters, this approach could be further extended to discriminate the OAM information multiplexed in multiple wavelengths and polarizations.

## Introduction

In 1992, Allen *et al.* proposed the concept of the orbital angular momentum (OAM) of light[1]. The typical example is a Laguerre-Gaussian (LG) beam with an azimuthal phase distribution of $e^{-il\varphi}$ which carries an OAM of $l\hbar$ per photon. Here, $\varphi$ is the angular coordinate in the transverse plane and $l$ is the so-called topological charge (TC). In the past two decades, optical OAM beams have been widely used in optical tweezers, information processing, imaging, and nonlinear optics[2-8]. Recently, scientists have found that OAM can provide an additional degree of freedom for optical communications[9,10]. By employing OAM multiplexing, one can greatly enhance the channel capacity and the spectral efficiency. The first experimental breakthrough was a terabit free-space data transmission in 2012[10]. Later, terabit-scale OAM mode division multiplexing was realized in a specially designed multimode fiber (with a typical core diameter of 15 μm)[9], which opens a door to apply OAM in high-capacity optical fiber networks[11]. Since then, significant advances have been achieved in fabricating integrated devices to manipulate the OAM states for different applications[12-16]. Researchers have manufactured compact OAM emitters by using, for example, an angular grating with a radius of 3.9 μm, a microstructure optical fiber tip of 4 μm in radius, and a microring resonator with a radius of 4.5 um[12,14,16]. Recently, the emitter size has been decreased further down to ~1 μm in radius by using a tunable open microcavity to produce various OAM modes[17]. The on-chip OAM multiplexing of broadband light was realized by using a mode-sorting nanoring aperture with a scale of 4.2 μm×4.2 μm[18]. However, it is still difficult to easily detect the OAM mode in an integrated device, especially when the OAM-carrying beam has a size of few microns.

In free space, there are many methods to detect an OAM state, which can be generally divided into two main categories. The traditional and efficient way is to build an interferometer[19]. By interfering an OAM beam with its own mirror image or a reference light beam with a known profile, the TC can be counted from the interference fringes[19-21]. The other category is to introduce mode converters, such as single slit[22], double slits[23], gradually–changing–period grating[24], triangular apertures[25], special designed refractive element[26], cylindrical lens[27] and fork grating[6], where the OAM modes can be analyzed from the converted patterns. So far, only a few nanophotonic devices including holographic coupler and plasmonic metasurface can realize OAM measurements on chip[28-31]. These reported devices have typical sizes of tens of microns, which need to be further reduced to match the compact OAM emitters[12] or the OAM fibers[9] for practical applications. In addition, precision nano-fabrication and careful optical alignment are generally required in such on-chip detectors. With few exceptions like weak measurement[32], the OAM mode is usually destroyed in most of the above detection schemes, which cannot be used to inspect the OAM mode in situ.

In this Letter, we demonstrate the use of a single plasmonic nanohole to measure the OAM state nondestructively. Such nanohole is a fundamental nanophotonic block, which can control the light beyond the diffraction limit through its coupling with plasmons[33]. The advantages of our method can be summarized in three folds. (1) The device size is reduced down to nanoscale, which can be easily installed on a fiber tip or an integrated OAM emitter. (2) The OAM information is extracted in a nondestructive way. In general, a plasmonic device suffers from significant loss, which originates from the substantial reflection at the metal surface and the inevitable increase of light absorption and scattering in the nanostructures. In our nanohole device, only a small portion of light is transmitted for the OAM detection. The high reflectivity (>94% in our current device) at the metal surface guarantees that the major OAM-carrying beam can still be applied for further tasks. The increase of light absorption and scattering due to the single nanohole structure is basically negligible. (3) Such metallic nanohole device can be easily fabricated with low-cost, and is flexible and robust for on-chip applications. It is particularly useful to monitor the OAM mode in situ in optical fiber networks or long-distance optical communications.

**Principle**

The device is a single nano-sized hole fabricated in a thin metal film. As shown in Fig. 1(a), the working principle of the device can be considered as a self-referenced interferometer. A light beam carrying an OAM of $l$ is normally incident onto the device. Because of the high reflectivity at the metal surface, most of the incident OAM beam is reflected back without significant change (See Supplementary Fig. 2). The light passing through the device takes two channels. One is an efficient way: the light couples with plasmons in the nanohole, travels through it, and emits photons at the other side. However, only the minority of the incident beam near the nanohole could take this route. The majority part of the transmitted beam goes directly through the metal film unaffected by the nanohole. Because of the skin depth of metal, the directly transmitted beam is attenuated by orders of magnitude, but still carries the original OAM information. By analyzing the interference pattern between the lights transmitted through these two channels, one can well investigate the property of the incident OAM beam.

Because of the strong confinement of light, the nanohole can be approximately seen as a point source. Therefore, the interference pattern can be considered as the superposition of an OAM mode and a spherical wave. As shown in Figs. 1(b) to 1(d), the generated pattern is a spiral structure or a distorted fork grating, depending on the position of the nanohole relative to the OAM beam profile. The TC of the OAM mode can be determined by counting the fringes of the interference pattern. For example, when an OAM beam with $l = 8$ coaxially interferes with the spherical wave, i.e., the nanohole is aligned with the center of the incident beam, the generated pattern has a spiral structure as shown in Fig. 1(b). The observed eight fringes, starting from the central singularity and winding around it, indicate the original TC of the beam. The sign of the TC can also be distinguished from the winding direction of fringes, i.e. clockwise (or anti-clockwise) winding represents a positive (or negative) TC. If the nanohole slightly deviates from the beam center or tilts a little, the spiral pattern becomes deformed as shown in Fig. 1(c). However, the number of fringes remains to be $l$ and can still be easily counted. When the deviation is severe, the interference pattern turns into a distorted fork grating as shown in Fig. 1(d). In this case, one should obtain the TC by calculating the difference between the numbers of fringes on the two sides of the fork. For example, in Fig. 1(d), the TC is $l = 15 - 7 = 8$.

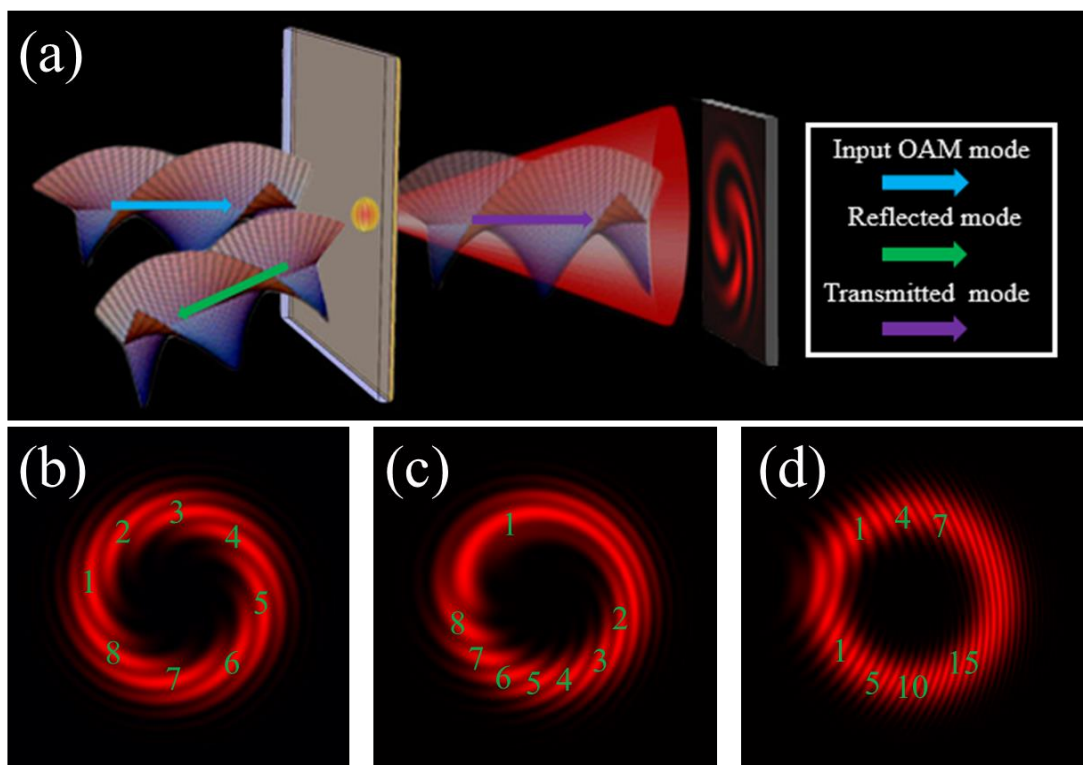

Figure 1. (a) The schematic diagram of the detecting process. An OAM beam is incident onto the nanohole device fabricated on an Au film. Most of the input beam is reflected for other uses. The spherical wave emitted from the nanohole interferes with the directly transmitted OAM mode. The TC can be determined from coaxial interference pattern (b), slightly off-axis interference pattern (c), or severe off-axis interference pattern (d) (see **Methods** for more information about the simulation).

The design of the device is to realize a good interference contrast at certain imaging plane. To achieve that, one must balance the directly-transmitted OAM mode and the spherical wave emitted from the nanohole channel so that they have similar power densities at the imaging plane. This can be achieved by optimizing the size of the nanohole and the thickness of the metallic film. The hole in the metal film is required to be small enough so that 1) it can work as a point source for spherical wave and 2) its disturbance to the transmitted and reflected OAM modes can be basically ignored. Considering that the OAM mode in an integrated device has a typical size of a few microns, the hole size should be no bigger than a few hundreds of nanometers. When light passes through such a nanohole, the transmissivity is greatly enhanced by plasmons. There are resonant peaks at certain wavelengths, i.e., enhanced optical transmission[33], because of the excitation of the localized surface plasmons (LSPs). For example, the Au nanohole of 300 nm in diameter in our experiment has a resonant peak at a wavelength of 633 nm (Fig. 2(a)). Since the transmissivity of the metal film can be estimated by the skin depth, one can design the nanohole device based on above discussions. Moreover, precision fabrication of the nanohole device is not strictly required. The robustness of our device partially lies in the fact that even when the nanohole is not perfectly fabricated, one can still optimize the quality of the interference pattern by properly selecting the observation plane. Because the spherical wave emitted from the nanohole diffracts much faster than the OAM beam, the interference contrast can be continuously tuned by moving the imaging plane.

**Experimental demonstrations**

**Sample fabrication.** One can have flexible platforms to support such metallic nanohole devices, depending on the working circumstance. For example, it can be installed on a fiber tip for in-situ OAM inspection in fiber networks, or set at the output port of a micro-emitter for mode analysis. In our experimental demonstration, the nanohole device is fabricated in a 230 nm-thick Au film on a glass substrate by a focused ion beam. The Au film has a transmissivity of ~4×10[-5] and a reflectivity of 94%. The loss of ~6% is mainly caused by the absorption and scattering in the Au film, which can be further suppressed by improving the film quality. Five nanohole devices with diameters of 250 nm, 300 nm, 350 nm, 400 nm and 450 nm are manufactured for comparisons. The inset in Fig. 2(a) shows the SEM picture of the 300 nm nanohole. The transmissions of the nanohole at an input wavelength of 633 nm are roughly measured by using a CCD camera. The relative transmission efficiency of the nanoholes (normalized to their areas) are shown in Fig. 2(a). Clearly, a resonant peak for the 633 nm laser appears at the 300 nm nanohole.

**The evolution of interference fields.** The performance of the single nanohole device is tested by using a He-Ne laser. The 633 nm laser passes through a vortex phase plate (VPP) to produce an OAM beam as the input light source. The beam size is ~15 μm in diameter, which is close to the OAM mode in a fiber[9]. In principle, the single nanohole can measure OAM beams with different TC values. Here, we choose $l = 8$ as an example. See **Method** and Supplementary Fig. 1 for detail information about the optical setup. To better utilize the plasmonic resonance, we use the 300 nm nanohole for the test. Since the center of an ideal OAM-carrying beam is a phase singularity with zero intensity, the nanohole is set slightly off center to improve the interference contrast. Therefore, a distorted spiral structure is expected in the transmitted interference pattern. The reflected beam well preserves the input OAM mode (see Supplementary Fig. 2). Figures 2(b)-2(d) show the evolution of the transmitted pattern along the propagation direction of light. The observation plane at a propagation distance of 90 um (Fig. 2(b)) presents a good image quality. One can observe eight curves winding around the central singularity, which is well consistent with the input TC of $l = 8$. As the propagation distance increases, the spherical wave diverges much more quickly than the vortex beam, which results in the reduction of the image quality as shown in Figs. 2(c) and 2(d). The contrast of the interference pattern deteriorates when the beam travels 120 um from the nanohole (Fig. 2(c)). At a distance of 160 um, the tails of the spiral curves are less distinguishable. The center of the interference field evolves into a pattern consisting of eight petals (Fig. 2(d)). Nevertheless, one can still determine the TC of the beam at a large range of observation planes. Such flexible working distance can be an important advantage in certain extreme operation conditions. The numerical simulations in Figs. 2(e)-(g) are well consistent with the experimental results.

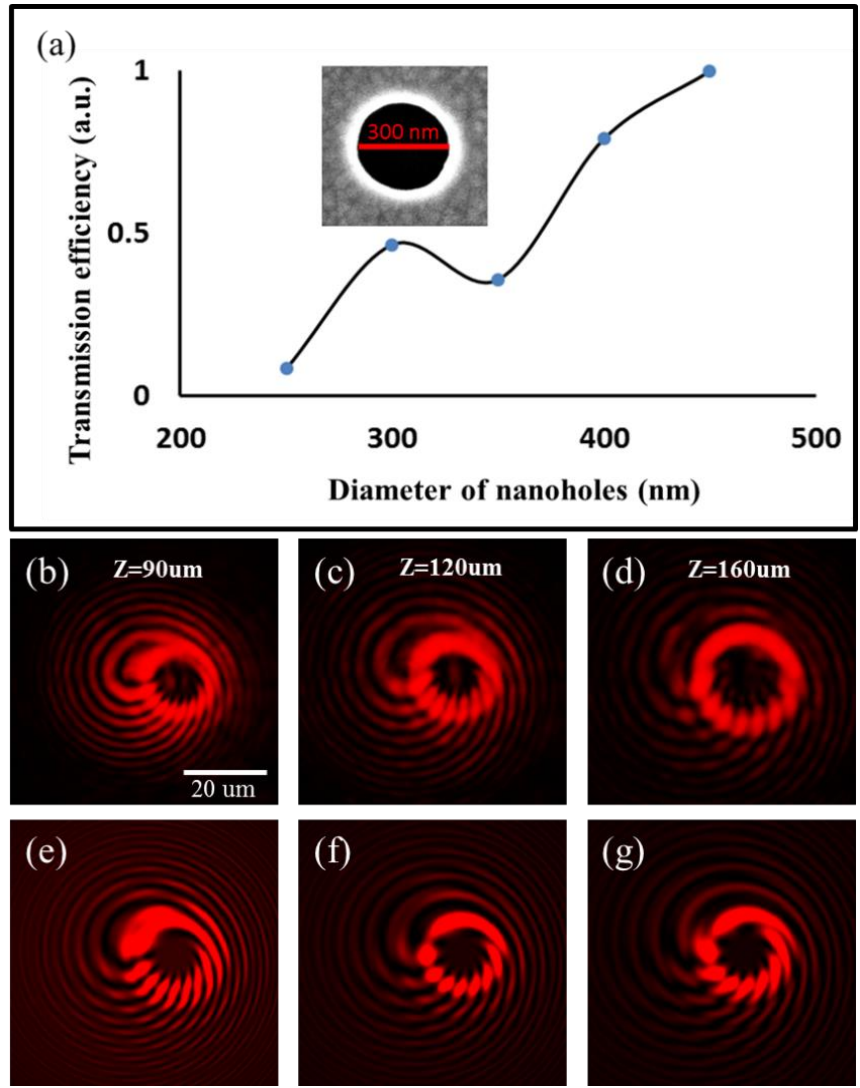

Figure 2. (a) The transmission efficiency of the nanoholes with various diameters normalized to their respective hole sizes. The inset is the SEM picture of a 300 nm-in-diameter nanohole. (b)-(d) show the evolution of the interference patterns when an OAM mode of $l = 8$ passes through the 300 nm nanohole device, which correspond to propagation distances of 90 μm, 120 μm, and 160 μm, respectively. The corresponding simulations are shown in (e)-(g), respectively. (Scale bar, 20 μm)

**Robustness tests**

In the above experiment, we have demonstrated that the image quality can be optimized by moving the observation plane. In real applications, there could be other unexpected errors. For example, the imprecision in nano-fabrication can result in inaccurate parameters of the nanohole. A deviation from the ideal position may occur when one installs the nanohole device. Therefore, we have run more tests to show the robustness of such single nanohole device in the presence of an error.

**Influence of the error in the nanohole size.** Figure 3 shows the interference patterns of different nanohole devices under the same experimental conditions. The observation plane is set at 100 μm away from the nanohole. The patterns in Figs. 3(a)-(d) correspond to the nanoholes with diameters of 250 nm, 300 nm, 350 nm, and 400 nm, respectively. In the experiment, the nanohole is placed on the brightest point of the OAM beam, which ensures enough light intensity of the spherical wave for each nanohole device. The image at the output face of the Au film (see the insets in Fig. 3) indicates the position of the nanohole relative to the OAM beam. One can barely resolve eight pedals in Fig. 3(a). The low contrast is caused by the low intensity of the spherical wave from the 250 nm hole (see Fig. 2(a)). The image qualities of Figs. 3(b) and 3(c) are significantly improved in comparison to Fig. 3(a). Because of the plasmonic resonance, the transmission efficiency of the 300 nm hole is greatly enhanced to be comparable to the 350 nm hole (Fig. 2(a)). Therefore, the image contrasts in Figs. 3(b) and 3(c) are similar. When further increasing the nanohole size to 400 nm, more detail features appear at the outer area of the pattern because of the strong spherical wave (Fig. 3(d)). Our results show that the performance of the nanohole device has a large tolerance for the size uncertainty. In addition, plasmons play a much more important role in smaller nanohole because the transmission efficiency under non-resonant condition falls very quickly when scaling down the size of the nanohole[34].

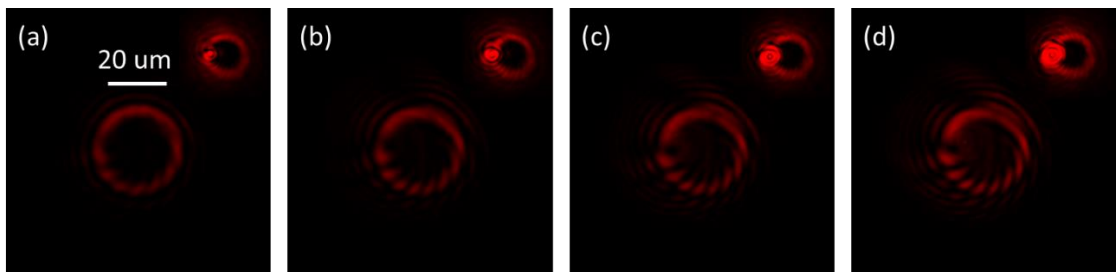

Figure 3. Comparison of the interference patterns when using nanoholes of (a) 250 nm, (b) 300 nm, (c) 350 nm, and (d) 400 nm in diameter, respectively. The imaging plane is set at a distance of 100 μm away from the sample. Each inset shows the image at the output face of the nanohole device, which is used to determine the position of the nanoholes relative to the OAM beam.

**Influence of the position error in the device installation.** Such error may often happen in practical installation and operation because of the nanoscale device, which is equivalent to the deviation of the nanohole location relative to the OAM beam. As discussed in the **Principle** part, this may modify the transmitted interference pattern because of the change in the relative position between the OAM mode and the spherical wave. However, the TC of the OAM beam can still be easily counted from the interference pattern. A nanohole with a diameter of 400 nm is used for the test. It is first set near the central singularity point of the input OAM beam, i.e., under a nearly coaxial-interference configuration (Fig. 4(a)). A clear spiral pattern is formed at the imaging plane of 60 μm away (Fig. 4(b)). Since the intensity close to the singularity of the OAM mode is nearly zero, the image contrast is relatively poor. As a comparison, the nanohole is moved near the bright ring (Fig. 4(c)). The spiral pattern is distorted as predicted in the **Principle** (Fig. 4(d)). However, the image contrast is greatly enhanced. Our device shows robustness to a relative displacement of the nanohole position. It should be mentioned that the decrease in the interference quality due to above errors could be partially compensated by shifting the imaging plane.

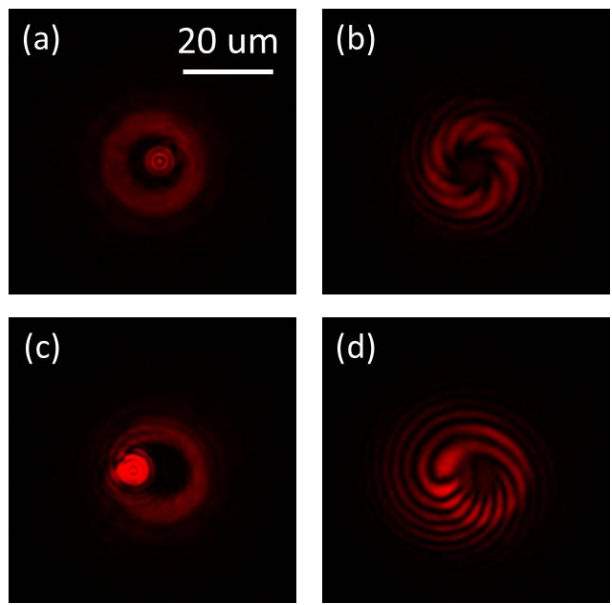

Figure 4. Influence of the nanohole position relative to the OAM beam on the interference pattern. (a) and (c) show the images at the output face of the nanohole device, which indicate the relative positions of the nanoholes. When the nanohole is placed near the OAM beam center (a), the interference pattern shows a clear spiral structure (b) with a low contrast at the observation plane of 60 μm. When the nanohole is set on the bright ring (c), the spiral pattern is distorted as shown in (d). The image quality is significantly improved due to the better matched interfering intensities.

## Discussion

We have experimentally demonstrated a simple scheme of using a single plasmonic nanohole to detect the OAM state in a nondestructive way. Due to greatly enhanced transmission by plasmons, single nanohole can effectively emit a spherical wave, which serves as a reference beam to detect the OAM mode through their interference. Since the nanohole size can be fabricated beyond the diffraction limit of light, it can easily operate on various integrated platforms to detect different spatial optical beams. Comparing to the traditional OAM detectors, our nanohole device has unique advantages in easy fabrication, low cost, nondestructiveness, and system integration. Its robustness under various operation conditions also makes itself a favorable choice for in-situ OAM mode monitoring. Such nanohole device can be further improved by growing high-quality metal thin films to reduce the losses of photons and plasmons during their propagations. In principle, the nanohole device can have a flexible size, depending on the spatial beam to be measured. The applications of such nanohole device are beyond the OAM detection. Since its working principle is a self-referenced interferometer, the single nanohole can also be used to measure the spatial phase distributions of, for example, Bessel beam, spherical wave, and high-order LG modes (see Supplementary Fig. 3) in integrated platforms. The excitation of plasmons is sensitive to wavelength as demonstrated in Fig. 2(a), and polarization if fabricating a non-symmetric hole[35,36], which can be utilized for discriminating the OAM multiplexed in multiple wavelengths and polarizations in optical fiber communications[18,30,37]. In addition, the device can be extended to mid-infrared and terahertz frequencies by fabricating nanoholes on alternative plasmonic materials such as metal alloys, highly-doped semiconductors, and graphene[38-41].

## Methods

**Experimental setup.** As shown in Supplementary Fig. 1, a He-Ne laser light source generates a linearly-polarized Gaussian beam at the wavelength of 633 nm, which passes through a VPP with a TC of 8 to form an OAM mode. A 20X objective after the VPP is used to focus the OAM beam with a diameter of ~15 μm, comparable to a typical fiber OAM mode. The reflected beam from the nanohole device is analyzed by a standard interferometer (Supplementary Fig. 2), which well preserves its original OAM information. The transmission signal is collected by a 10X objective and imaged on a CCD camera for mode analysis. The observation plane is selected by carefully moving both the10X objective and the CCD camera.

**Numerical simulations.** To verify the measured results, numerical calculations by Matlab programming are carried out to simulate the detection process. In the simulations, the input beam propagates along the z axis. The nanohole device is located at z = 0. We define the input OAM mode as $u_I = u_{OAM}\left(x_1, y_1\right)$. The simplified transmission function $T\left(x_1, y_1\right)$ of the nanohole device can be expressed as

$$T\left(x_1, y_1\right) = \begin{cases} t_{film}, & \sqrt{\left(x_1 - X\right)^2 + \left(y_1 - Y\right)^2} \geq R \\ 1, & \sqrt{\left(x_1 - X\right)^2 + \left(y_1 - Y\right)^2} < R \end{cases}, \quad (1)$$

where $t_{film}$ is the transmittivity of the metallic film, X and Y are the coordinates of the nanohole with a radius of R. The output field at z = 0 from the nanohole device can be written as

$$u_o\left(x_1, y_1, z = 0\right) = u_{OAM}\left(x_1, y_1\right) T\left(x_1, y_1\right). \quad (2)$$

Using the Fresnel approximation, the output field in the detection plane at a distance of $z$ is

$$u_o\left(x_2, y_2, z\right) = \iint u_o\left(x_1, y_1, z = 0\right) h\left(x_2 - x_1, y_2 - y_1, z\right) dx_1 dy_1, \quad (3)$$

where $h\left(x, y, z\right)$ is defined by

$$h\left(x, y, z\right) = \frac{e^{ikz}}{i\lambda z} e^{i\frac{k}{2z}\left(x^2+y^2\right)}. \qquad (4)$$

By substituting Eq. (2) into Eq. (3), one can obtain

$$u_o\left(x_2, y_2, z\right) = \iint u_{OAM}\left(x_1, y_1\right) T\left(x_1, y_1\right) h\left(x_2 - x_1, y_2 - y_1, z\right) dx_1 dy_1. \qquad (5)$$

If $R$ is much smaller than the size of input OAM mode, the transmission function in Eq. (1) can be further simplified as

$$T\left(x_1, y_1\right) = t_{film} + A\delta(X, Y), \qquad (6)$$

where A is the amplitude modulation factor. When applying this approximation, Eq. (5) becomes

$$u_o\left(x_2, y_2, z\right) = t_{film}\iint u_{OAM}\left(x_1, y_1\right) h\left(x_2 - x_1, y_2 - y_1, z\right) dx_1 dy_1 \\ + Ah\left(x_2 - X, y_2 - Y, z\right). \qquad (7)$$

Equation (7) indicates that the field in the detection plane can be seen as the interference between an OAM mode and a spherical wave. The detected intensity interference pattern on the CCD camera is

$$I = \left|u_o\right|^2. \qquad (8)$$

By changing z in the transfer function, the interference results can be obtained at different propagation distances. Supplementary Fig. 4 shows the simulated results of a 2 um-in-diameter hole, which are well consistent with the experimental patterns.

## Acknowledgement

This work was supported by National Basic Research Program of China (No. 2012CB921804), National Science Foundation of China (Nos. 11274162, 11404165, and 11321063), and Priority Academic Program Development of Jiangsu Higher Education Institutions (PAPD).

## Author contributions

D.W., Y.W., D.L., Y.Z., W. Z. and X. F. performed the experiments and simulations under the guidance of Y.Z. and M.X. D.W., Y.Z., and M.X. wrote the manuscript with contributions from all co-authors.

## References

1    Allen, L., Beijersbergen, M. W., Spreeuw, R. J. & Woerdman, J. P. Orbital angular momentum of light and the transformation of Laguerre-Gaussian laser modes. *Phys. Rev. A* **45**, 8185-8189 (1992).

2    Grier, D. G. A revolution in optical manipulation. *Nature* **424**, 810-816 (2003).

3    Simpson, N. B., Dholakia, K., Allen, L. & Padgett, M. J. Mechanical equivalence of spin and orbital angular momentum of light: an optical spanner. *Opt. Lett.* **22**, 52-54 (1997).

4    Nicolas, A. *et al.* A quantum memory for orbital angular momentum photonic qubits. *Nature Photonics* **8**, 234-238 (2014).

5    Bloch, N. V. *et al.* Twisting Light by Nonlinear Photonic Crystals. *Phy. Rev. Lett.* **108**, 233902 (2012).

6        Mair, A., Vaziri, A., Weihs, G. & Zeilinger, A. Entanglement of the orbital angular momentum states of photons. *Nature* **412**, 313-316 (2001).

7        Furhapter, S., Jesacher, A., Bernet, S. & Ritsch-Marte, M. Spiral phase contrast imaging in microscopy. *Opt. Express* **13**, 689-694 (2005).

8        Fang, X. *et al.* Coupled orbital angular momentum conversions in a quasi-periodically poled LiTaO3 crystal. *Opt. Lett.* **41**, 1169-1172 (2016).

9        Bozinovic, N. *et al.* Terabit-scale orbital angular momentum mode division multiplexing in fibers. *Science* **340**, 1545-1548 (2013).

10       Wang, J. *et al.* Terabit free-space data transmission employing orbital angular momentum multiplexing. *Nature Photon.* **6**, 488-496 (2012).

11       Li, S., Mo, Q., Hu, X., Du, C. & Wang, J. Controllable all-fiber orbital angular momentum mode converter. *Opt. Lett.* **40**, 4376-4379 (2015).

12       Cai, X. *et al.* Integrated compact optical vortex beam emitters. *Science* **338**, 363-366 (2012).

13       Li, H. *et al.* Orbital angular momentum vertical-cavity surface-emitting lasers. *Optica* **2**, 547 (2015).

14       Rodrigues Ribeiro, R. S., Dahal, P., Guerreiro, A., Jorge, P. & Viegas, J. Optical fibers as beam shapers: from Gaussian beams to optical vortices. *Opt. Lett.* **41**, 2137-2140 (2016).

15       Brasselet, E., Murazawa, N., Misawa, H. & Juodkazis, S. Optical vortices from liquid crystal droplets. *Phys. Rev. Lett.* **103**, 103903 (2009).

16       Miao, P. *et al.* Orbital angular momentum microlaser. *Science* **353**, 464-467 (2016).

17       Dufferwiel, S. *et al.* Spin Textures of Exciton-Polaritons in a Tunable Microcavity with Large TE-TM Splitting. *Phys Rev Lett* **115**, 246401 (2015).

18       Ren, H., Li, X., Zhang, Q. & Gu, M. On-chip noninterference angular momentum multiplexing of broadband light. *Science* **352**, 805-809 (2016).

19       Padgett, M., Arlt, J., Simpson, N. & Allen, L. An experiment to observe the intensity and phase structure of Laguerre–Gaussian laser modes. *Am. J. Phy.* **64**, 77 (1996).

20       Leach, J., Padgett, M. J., Barnett, S. M., Franke-Arnold, S. & Courtial, J. Measuring the Orbital Angular Momentum of a Single Photon. *Phy. Rev. Lett.* **88**, 257901 (2002).

21       Huguenin, J. A., Coutinho dos Santos, B., dos Santos, P. A. & Khoury, A. Z. Topological defects in moire fringes with spiral zone plates. *J. Opt. Soc. Am. A* **20**, 1883-1889 (2003).

22       Ghai, D. P., Senthilkumaran, P. & Sirohi, R. S. Single-slit diffraction of an optical beam with phase singularity. *Opt. Lasers Eng.* **47**, 123-126 (2009).

23       Sztul, H. I. & Alfano, R. R. Double-slit interference with Laguerre-Gaussian beams. *Opt. Lett.* **31**, 999-1001 (2006).

24       Dai, K., Gao, C., Zhong, L., Na, Q. & Wang, Q. Measuring OAM states of light beams with gradually-changing-period gratings. *Opt. Lett.* **40**, 562-565 (2015).

25       Hickmann, J. M., Fonseca, E. J. S., Soares, W. C. & Chávez-Cerda, S. Unveiling a Truncated Optical Lattice Associated with a Triangular Aperture Using Light's Orbital Angular Momentum. *Phy. Rev. Lett.* **105**, 053904 (2010).

26       Mirhosseini, M., Malik, M., Shi, Z. & Boyd, R. W. Efficient separation of the orbital angular momentum eigenstates of light. *Nat. Commun.* **4**, 2781 (2013).

27       Denisenko, V. *et al.* Determination of topological charges of polychromatic optical vortices. *Opt. Express* **17**, 23374-23379 (2009).

28       Genevet, P., Lin, J., Kats, M. A. & Capasso, F. Holographic detection of the orbital angular

momentum of light with plasmonic photodiodes. *Nat. Commun.* **3**, 1278 (2012).

29  Liu, A. P. *et al.* Detecting orbital angular momentum through division-of-amplitude interference with a circular plasmonic lens. *Sci. Rep.* **3**, 2402 (2013).

30  Mei, S. *et al.* On-chip discrimination of orbital angular momentum of light with plasmonic nanoslits. *Nanoscale* **8**, 2227-2233 (2016).

31  Yang, K. *et al.* Wavelength-selective orbital angular momentum generation based on a plasmonic metasurface. *Nanoscale* **8**, 12267-12271 (2016).

32  Malik, M. *et al.* Direct measurement of a 27-dimensional orbital-angular-momentum state vector. *Nat. Commun.* **5**, 3115(2014).

33  Genet, C. & Ebbesen, T. W. Light in tiny holes. *Nature* **445**, 39-46 (2007).

34  Roberts, A. Electromagnetic theory of diffraction by a circular aperture in a thick, perfectly conducting screen. *J. Opt. Soc. Am. A* **4**, 1970-1983 (1987).

35  Degiron, A., Lezec, H. J., Yamamoto, N. & Ebbesen, T. W. Optical transmission properties of a single subwavelength aperture in a real metal. *Opt. Commun.* **239**, 61-66 (2004).

36  Lezec, H. J. *et al.* Beaming light from a subwavelength aperture. *Science* **297**, 820-822 (2002).

37  Wei, D. *et al.* Survival of the orbital angular momentum of light through an extraordinary optical transmission process in the paraxial approximation. *Opt. Express* **24**, 12007-12012 (2016).

38  Gong, C. & Leite, M. S. Noble Metal Alloys for Plasmonics. *ACS Photonics* **3**, 507-513 (2016).

39  Yu, N. *et al.* Designer spoof surface plasmon structures collimate terahertz laser beams. *Nature Mater.* **9** (2010).

40  Ju, L. *et al.* Graphene plasmonics for tunable terahertz metamaterials. *Nature Nanotech.* **6**, 630-634 (2011).

41  Koppens, F. H., Chang, D. E. & Garcia de Abajo, F. J. Graphene plasmonics: a platform for strong light-matter interactions. *Nano Lett.* **11**, 3370-3377 (2011).

# Supplementary Information

Dunzhao Wei, Yongmei Wang, Dongmei Liu, Yunzhi Zhu, Weihao Zhong, Xinyuan Fang, Yong Zhang,[*] and Min Xiao[‡]

E-mails: zhangyong@nju.edu.cn and mxiao@uark.edu

## 1. Experimental setup

The schematic diagram of the experimental setup is shown in Fig. S1. A He-Ne laser light source generates a linearly polarized Gaussian beam at the wavelength of 633 nm, which passes through a VPP with a TC of 8 to form an OAM mode. A 20X objective after the VPP is used to focus the OAM beam with a diameter of ~15 μm comparable to a fiber OAM mode. The reflected beam from the nanohole device is analyzed by a standard interferometer. The transmission signal is collected by a 10X objective and imaged on a CCD camera for mode analysis. The observation plane is selected by moving the 10X objective and the CCD camera.

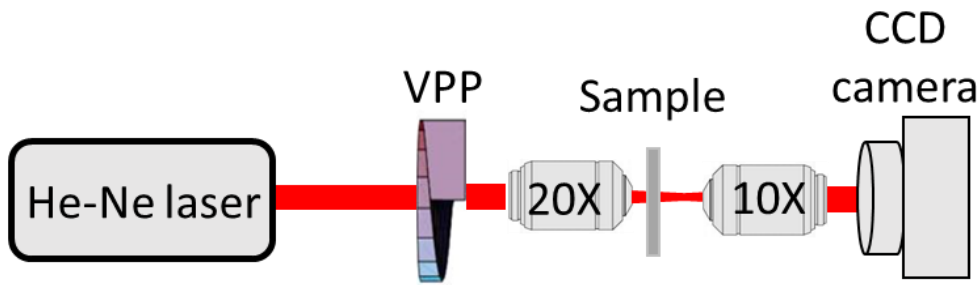

**Supplementary Fig. 1.** The schematic diagram of the experimental setup.

## 2. The characterization of the reflected OAM modes from various nanohole devices

The reflected OAM mode from the nanohole is measured by using a standard interference method[1]. The input OAM mode carries an OAM of $l = 8$. Eleven nanoholes with various diameters are tested. The reflected intensity pattern presents a well-defined ring shape. The interference pattern clearly shows an OAM of $l = 8$, which is well consistent with the input one (Fig. S2).

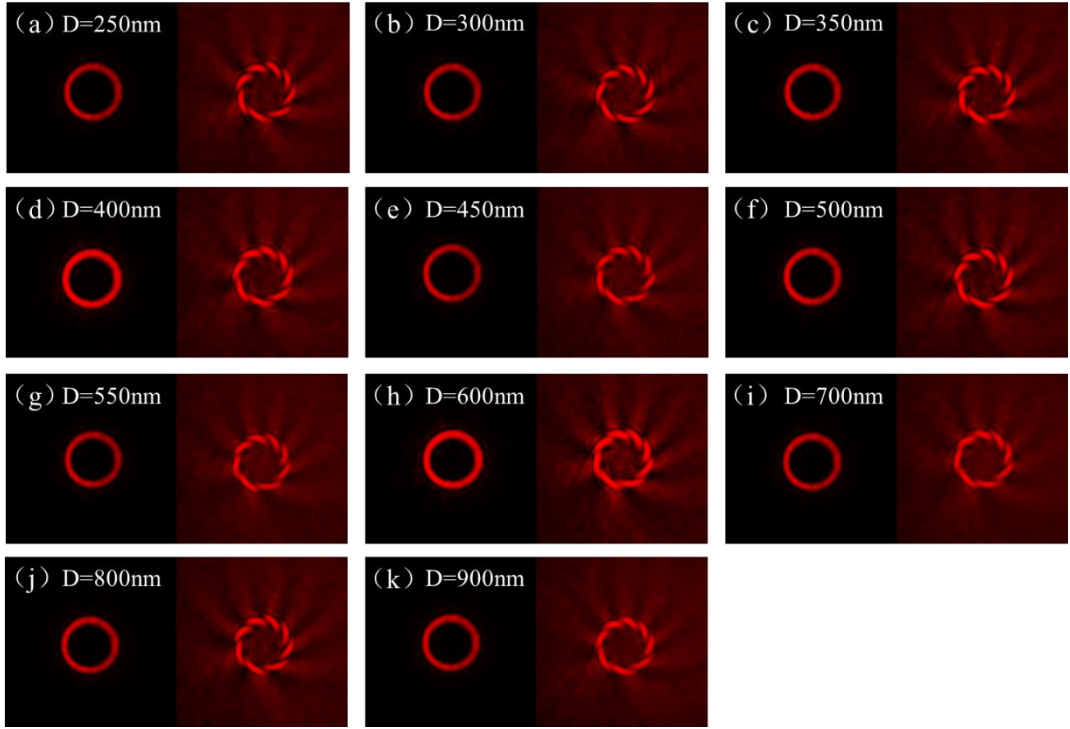

**Supplementary Fig. 2.** The reflected OAM beam from the nanohole devices with different diameters. In (a)-(k), the reflected intensity patterns (left) are recorded by a CCD camera. By using a standard interference method, the interference pattern (right) clearly shows an OAM of $l = 8$, which is well consistent with the input one. D is the diameter of the corresponding nanohole.

### 3. Measurement of various spatial lights using the nanohole device

In the experiment, a spatial light modulation (SLM) is used to generate different wavefronts. A He-Ne laser operating at 633 nm wavelength serves as the light source to match the working wavelength of the SLM. A 500 nm-in-diameter hole in a 0.4%-transmission Au film is used for demonstration. The nanohole is placed at the center of the incident pattern. In the experiment, we loaded various phase patterns on the SLM, including Bessel beam (Fig. S3(a)), sphere waves (Figs. S3(b) and S3(c)) and spiral waves (Figs. S3(d) and S3(e)). When these spatial light beams pass through the nanohole device, the corresponding interference patterns in Figs. S3(f)-(j) well present the loaded phase patterns.

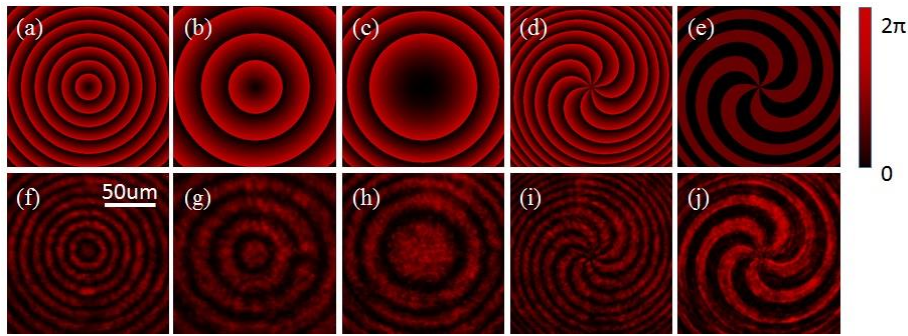

**Supplementary Fig. 3.** Detection of a spatial wavefront by using the single nanohole device. (a)-(e) are the phase patterns loaded on the SLM, which correspond to Bessel beam (a), sphere waves (b and c) and spiral waves (d and e), respectively. (f)-(j) are the corresponding interference patterns

when light beams pass through the nanohole device, which is placed at the center of the incident pattern.

## 4. Measured and calculated evolution of the interference intensity pattern after the device

The device used here is a 1 um-in-radius hole in a 0.4%-transmission Au film. The input light is an OAM beam with *l* = 8. The evolution of the interference pattern after the device is recorded by a CCD camera as shown in Fig. S4(a)-(e), which is well consistent with the calculated results in Fig. S4(f)-(j) by using Matlab programming.

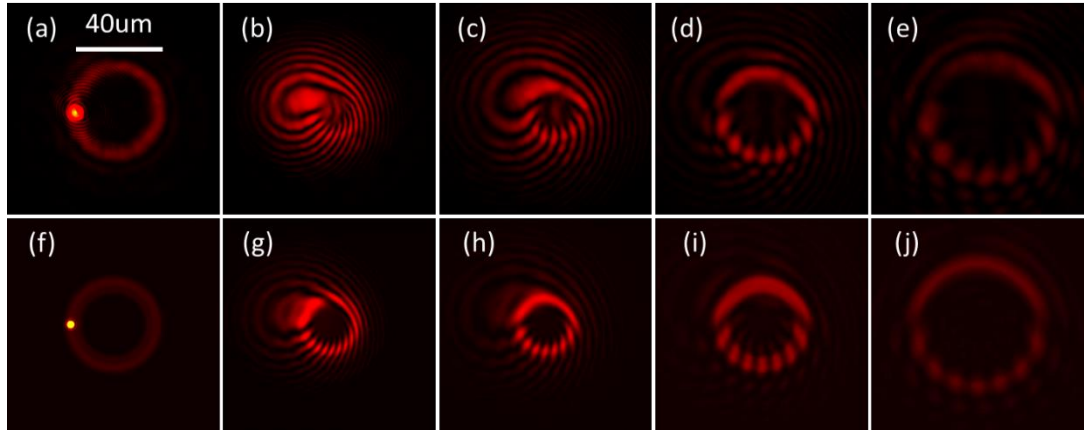

**Supplementary Fig. 4.** The comparison of interference patterns between experimental results and simulation results. (a)-(e) are the experimental results at the propagation distances of 0 um (i.e. the output face of the device), 100 um, 160 um, 260 um, and 400 um respectively. (f)-(j) are the simulation results by using Matlab programming, which are in good agreement with the corresponding experimental patterns.

## Reference

1    Padgett, M., Arlt, J., Simpson, N. & Allen, L. An experiment to observe the intensity and phase structure of Laguerre–Gaussian laser modes. *Am. J. Phy.* **64**, 77 (1996).